\documentclass[aps,prb,floatfix,showpacs,twocolumn]{revtex4}

\usepackage[dvips]{graphicx}

\begin{document}

\title{Scaling analysis of Schottky barriers at metal-embedded 
semiconducting carbon nanotube interfaces}



\author{ Yongqiang Xue $^{*}$ and Mark A. Ratner}
\affiliation{Department of Chemistry and Materials Research Center, 
Northwestern University, Evanston, IL 60208}
\date{\today}

\begin{abstract}
We present an atomistic self-consistent tight-binding study of the 
electronic and transport properties of metal-semiconducting carbon 
nanotube interfaces as a function of the nanotube channel length when 
the end of the nanotube wire is buried inside the electrodes. We show 
that the lineup of the nanotube band structure relative to the metal 
Fermi-level depends strongly on the metal work function but weakly on 
the details of the interface. We analyze the length-dependent transport 
characteristics, which predicts a transition from tunneling to 
thermally-activated transport with increasing nanotube channel length. 
\end{abstract}


\maketitle


The nature of Schottky barrier formation and its effect on charge transport 
through metal-semiconductor interfaces have been actively investigated for 
decades due to their importance in microelectronics technology, but are not 
fully resolved despite the tremendous efforts from both the experimental 
and theoretical sides.~\cite{MS} The rapid development of single-wall 
carbon nanotubes (SWNT) as a promising device technology functioning 
at the nano/molecular-scale~\cite{Dekker99,DM,Avouris} 
presents a new challenge.~\cite{Xue99,TersoffNT,OdinDe} 
SWNTs are nanometer-diameter all-carbon 
cylinders. Unlike the planar metal-semiconductor interface, both the contact 
area and the active device region in a metal-nanotube (NT) interface can 
have atomic-scale dimensions. In addition, due to the weaker effects 
of electron-impurity~\cite{Ando} and electron-phonon 
scattering~\cite{Phonon,NTDai} in quasi-one-dimensional systems, transport 
through nanotube junctions can be either coherent or phonon-limited 
depending on the nanotube type, channel length, temperature and the 
bias voltage. These have made it difficult to assess the Schottky 
barrier effect on the measured transport characteristics.  

\begin{figure}
\includegraphics[height=2.5in,width=2.8in]{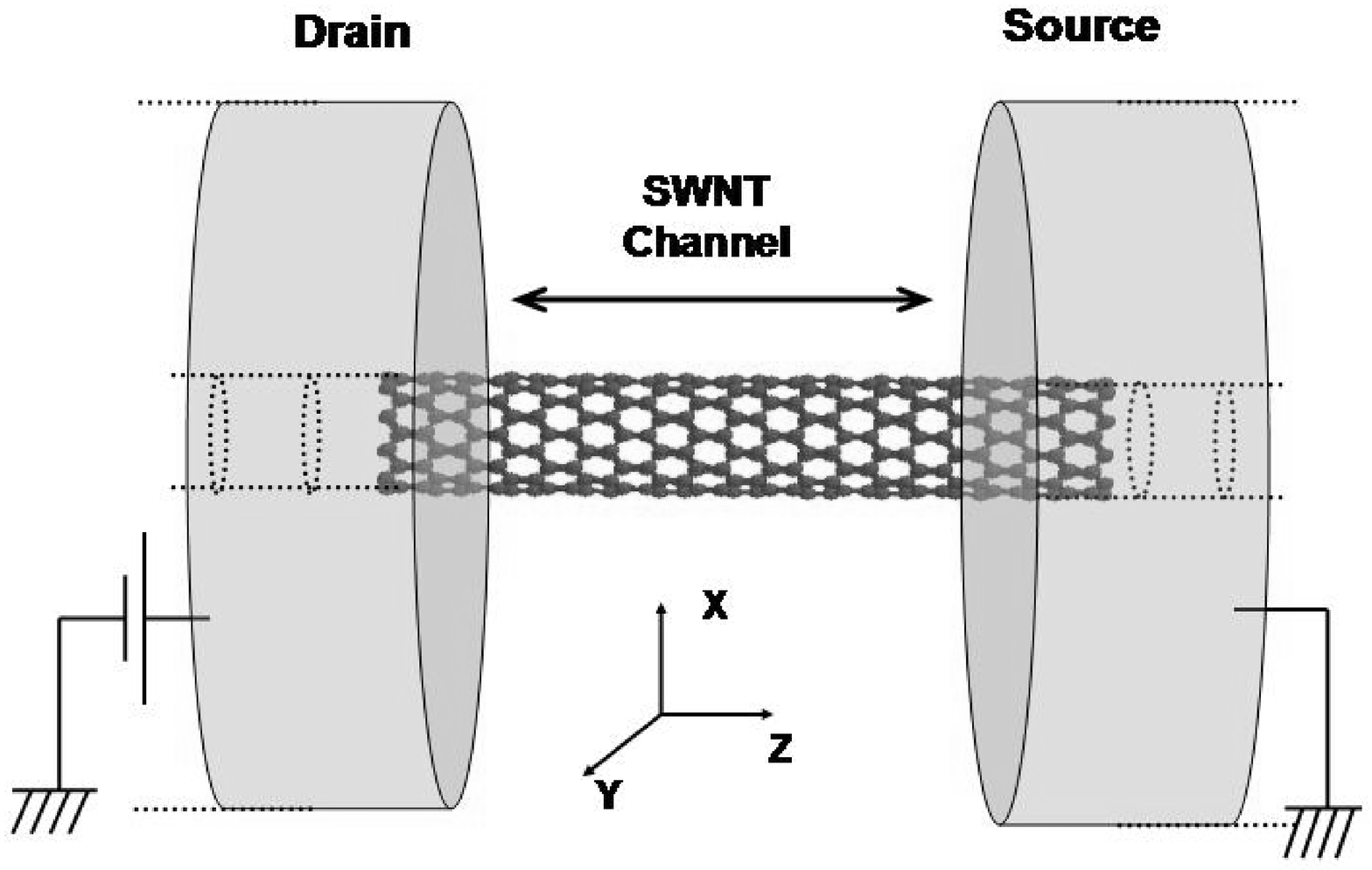}
\caption{\label{xueFig1} (Color online) Schematic illustration of the 
metal-SWNT wire-metal junction. The ends of the long SWNT wire are 
surrounded entirely by the semi-infinite electrodes, with only a finite 
segment being sandwiched between the electrodes (defined as the 
channel). Also shown is the coordinate system of the nanotube junction. }
\end{figure}

In this paper, we present an analysis of Schottky barrier formation at a 
model metal-SWNT interface as a function of the SWNT channel length. 
The model system is illustrated schematically in Fig.\ \ref{xueFig1}, where 
the ends of an infinitely long SWNT wire are buried inside the semi-inifinite 
metallic electrodes and the channel length is determined by the 
distance between the source and drain electrodes.~\cite{NTDai}  
The embedded contact scheme is favorable for the formation of 
low-resistance contact.~\cite{DM,Avouris,NTDai} We choose $(10,0)$ 
SWNT as the protype semiconducting SWNT, whose work function is taken 
as that of the graphite ($4.5$ eV).~\cite{DM} The SWNT channel length 
invesitigated ranges from $L=2.0,4.1,8.4,12.6,16.9$(nm) to $21.2$ (nm), 
corresponding to number of unitcells of $5,10,20,30,40$ and $50$ 
respectively. We calculate the transport characteristics within the coherent 
transport regime, as appropriate for such short 
nanotubes.~\cite{Phonon,NTDai} 
We consider gold (Au) and titanium (Ti) electrodes as examples of 
high- and low- work function metals ($5.1$ and $4.33$ eV respectively 
for polycrystalline materials). In view of recent report on ballistic transport 
in palladium (Pd) contacted SWNT devices,~\cite{DT03} we also consider 
Pd electrodes with a similar workfunction to gold ($5.12$ eV 
for polycrystalline materials).~\cite{CRC} 

We analyze the Schottky barrier effect at the metal-SWNT interface by 
examining the electrostatics, the band lineup and the conductance of 
the metal-SWNT wire-metal junction as a function of the SWNT channel 
length. This is investigated 
using a Green's function based self-consistent tight-binding (SCTB) theory, 
which takes fully into account the three-dimensional electrostatics and 
the atomic-scale electronic structure of the SWNT junctions. The SCTB 
model is essentially the semiempirical implementation of the 
self-consistent Matrix Green's function (SCMGF) method for \emph{ab initio} 
modeling of molecular-scale devices~\cite{XueMol} and has been 
used previously to investigate Schottky barrier formation when a finite-size  
SWNT molecule is contacted to the electrodes through the dangling 
bonds at the end.~\cite{XueNT} 

\begin{figure}
\includegraphics[height=2.0in,width=2.8in]{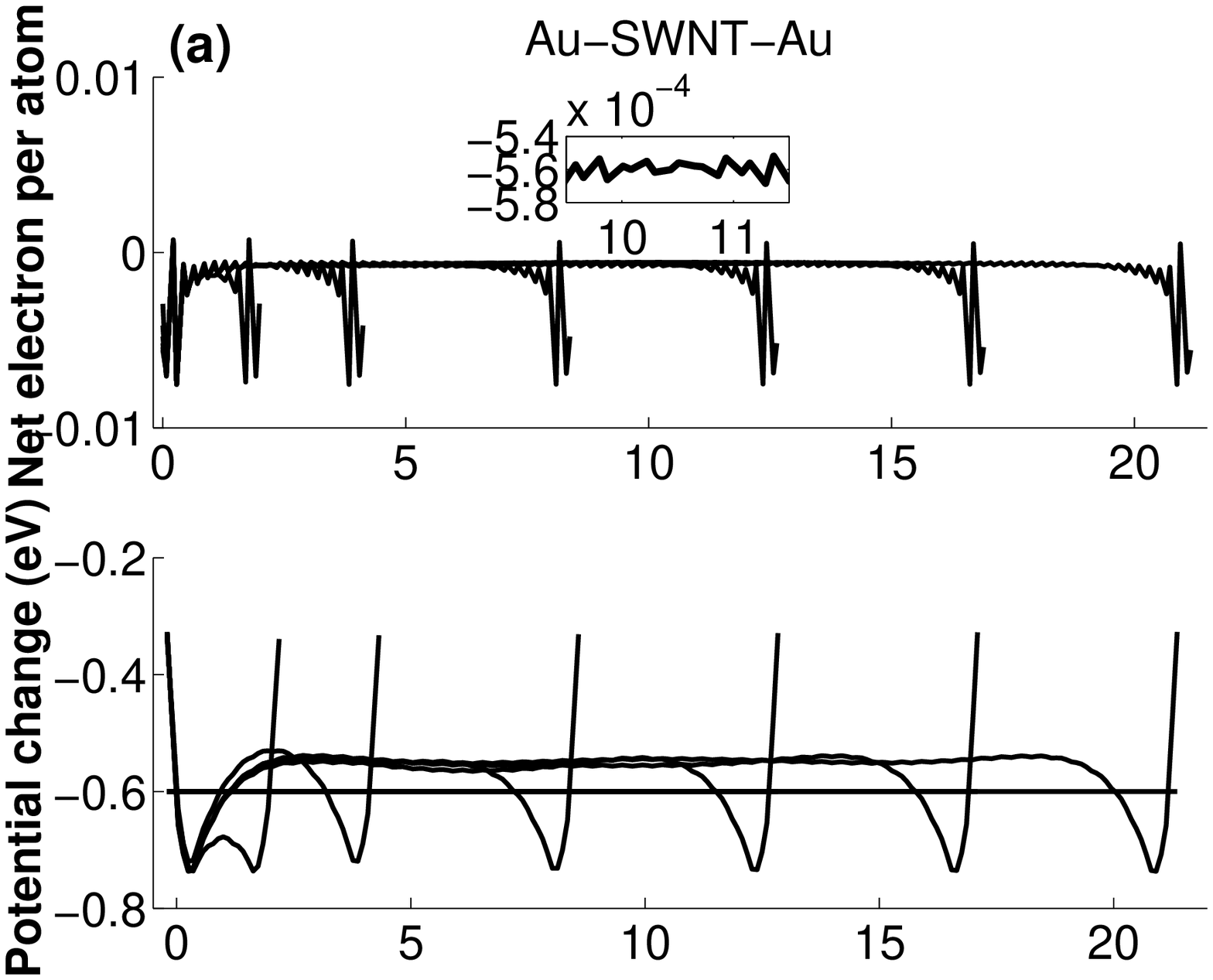}
\includegraphics[height=2.0in,width=2.8in]{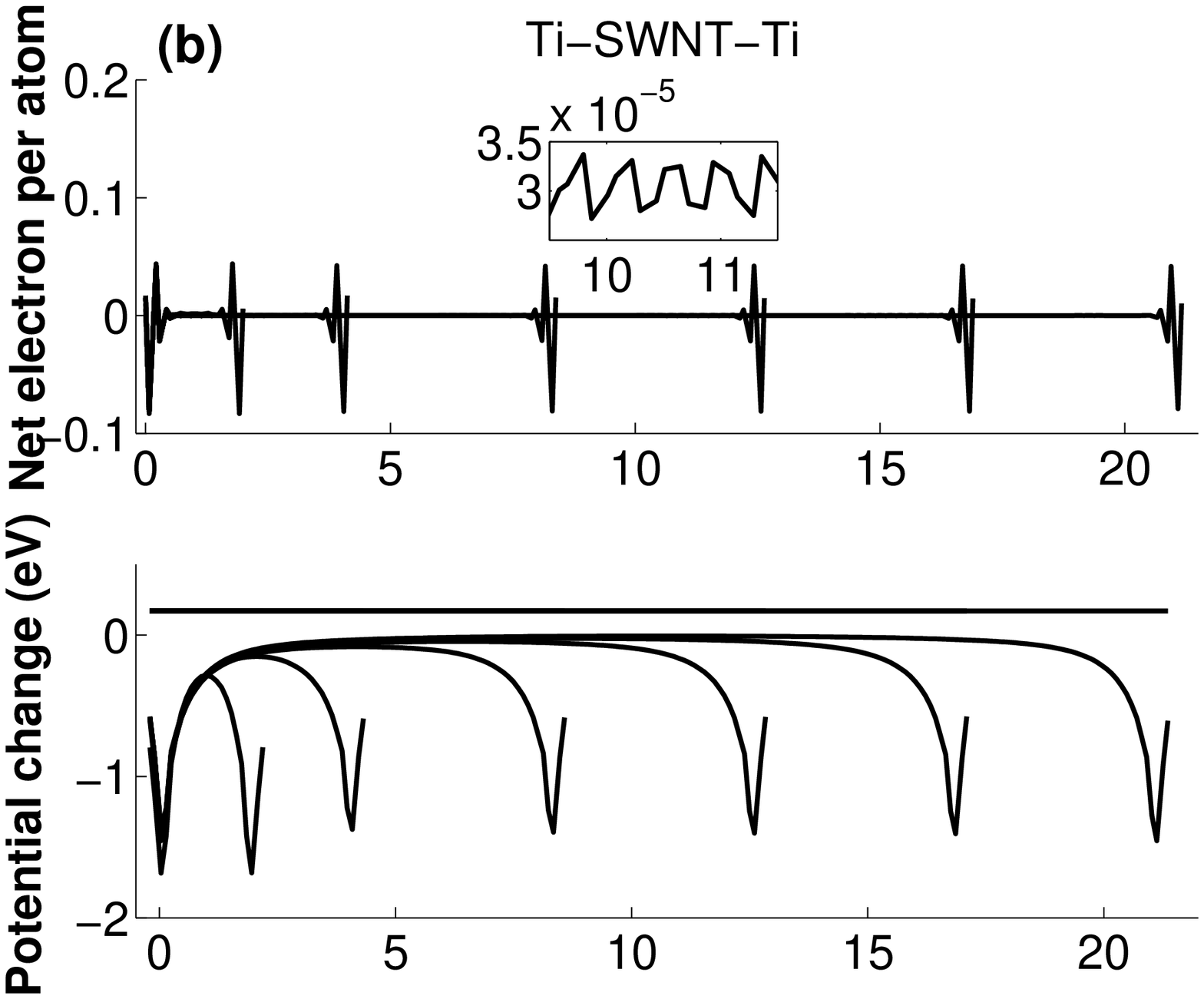}
\includegraphics[height=2.0in,width=2.8in]{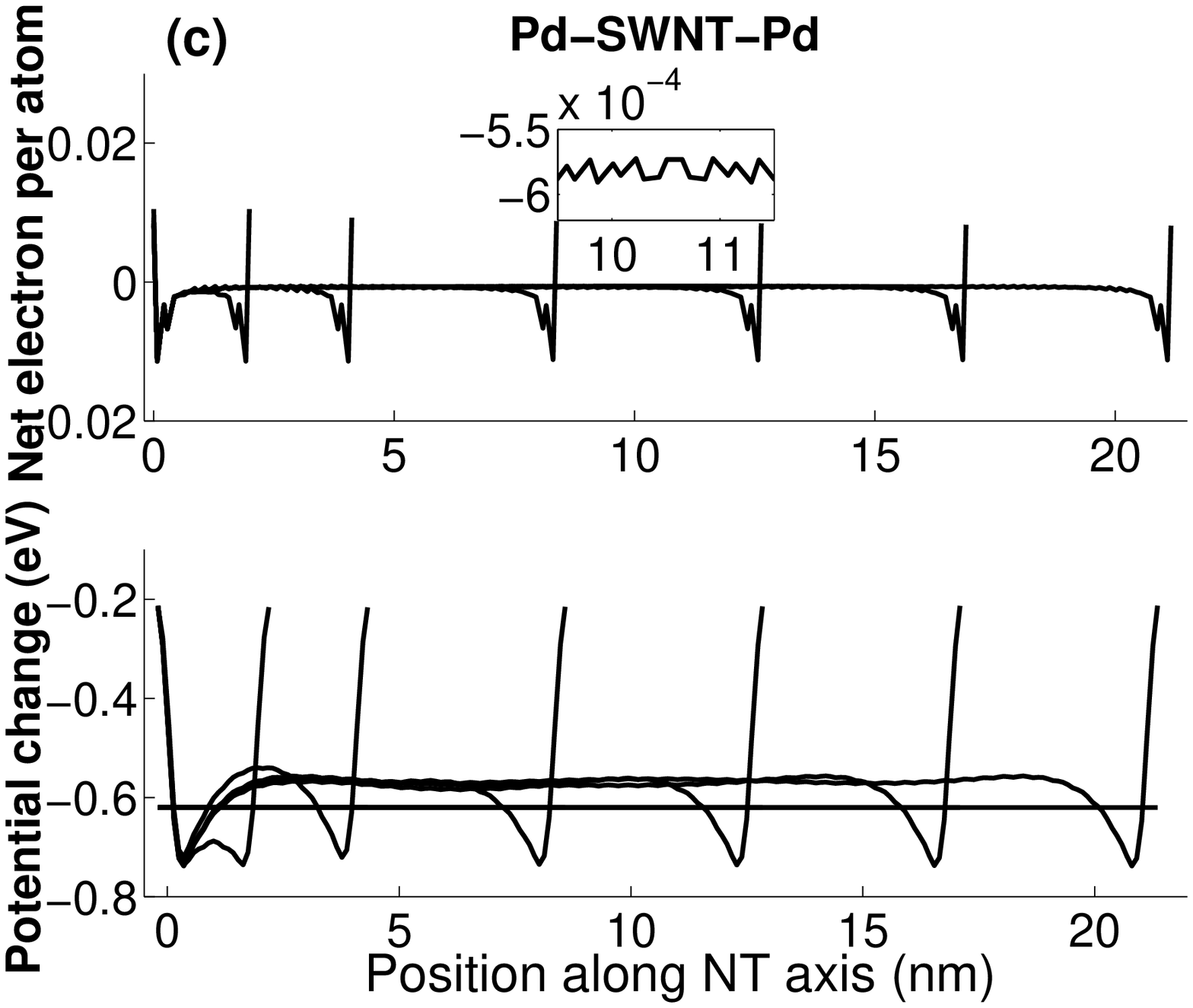}
\caption{\label{xueFig2} Electrostatics of the three metal-SWNT-metal 
junctions as a function of SWNT channel length for six different lengths. 
For each junction, the inset shows the magnified view of the transfered 
charge in the middle of the channel for the longest SWNT studied. 
The electrostatic potential change shown here is that taken at the 
cylindrical surface of the SWNT. The horizontal lines in the potential plot 
denote the work function differences between the electrodes 
and the SWNT wire.  }
\end{figure}

\begin{figure}
\includegraphics[height=2.5in,width=2.8in]{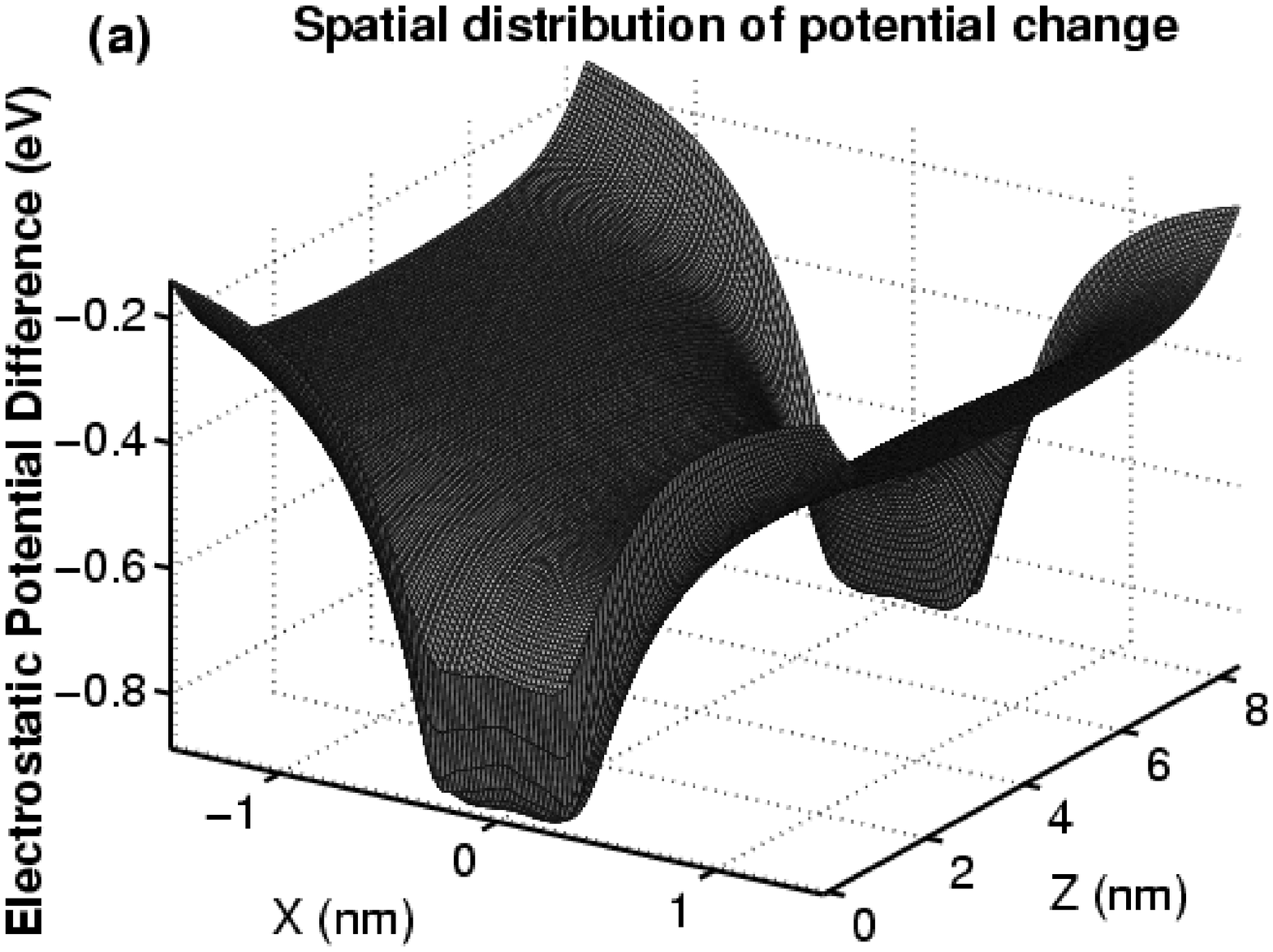}
\includegraphics[height=2.5in,width=2.8in]{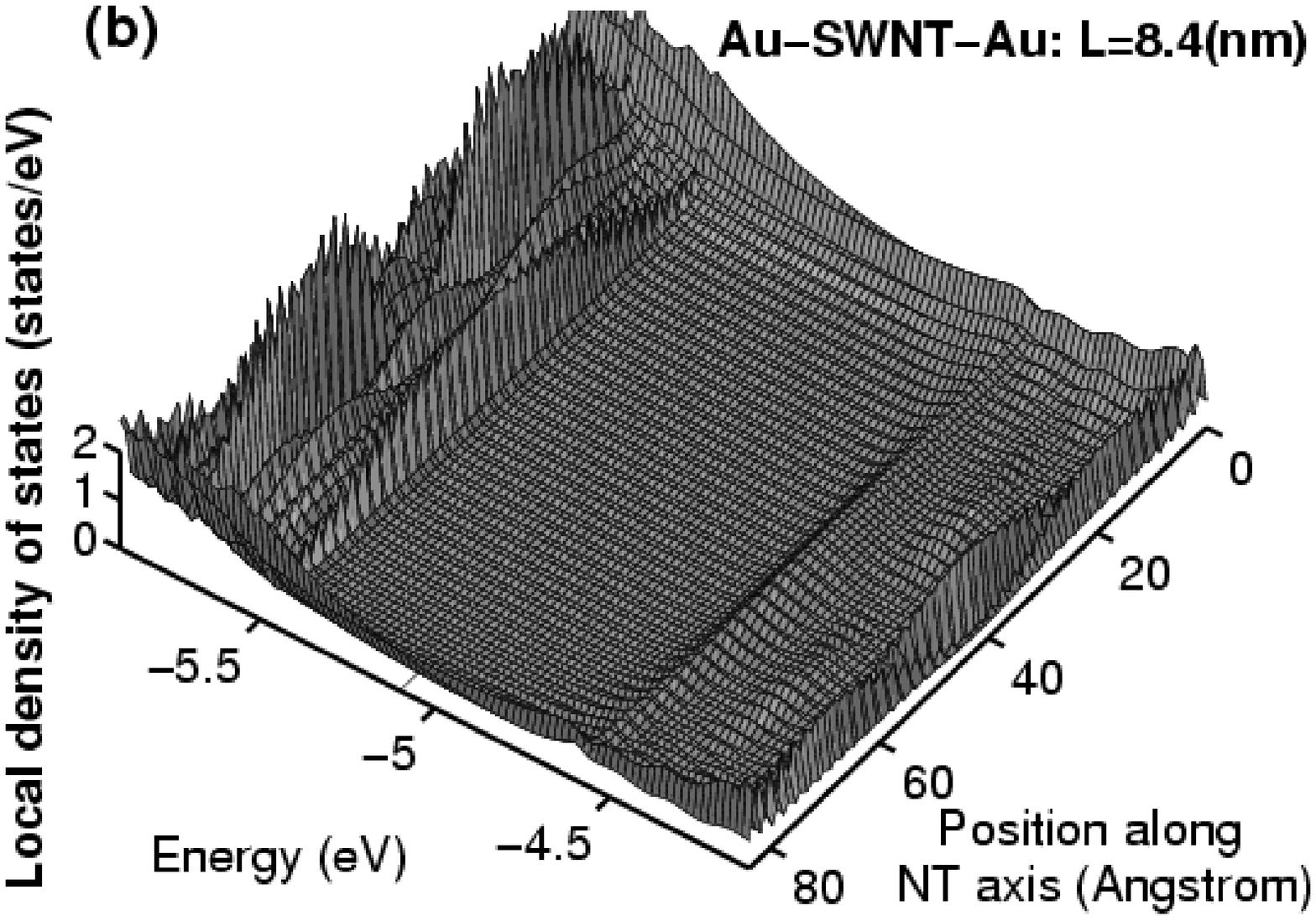}
\caption{\label{xueFig3} (Color online) Fig. (a) shows 
cross sectional view of electrostatic 
potential change at the Au-SWNT wire-Au junction for SWNT channel length 
of $8.4(nm)$. The SWNT diameter is $0.8(nm)$. (b) shows the 
corresponding local density of states (LDOS) as a function of position along 
the NT axis for SWNT channel length of $8.4(nm)$. The plotted LDOS is 
obtained by summing over the $10$ atoms of each carbon ring of the 
$(10,0)$ SWNT. Note that each cut along the energy axis at a given axial 
position gives the LDOS of the corresponding carbon ring and each cut along 
the position axis at a given energy gives the corresponing band shift. }
\end{figure}

\begin{figure}
\includegraphics[height=2.5in,width=2.8in]{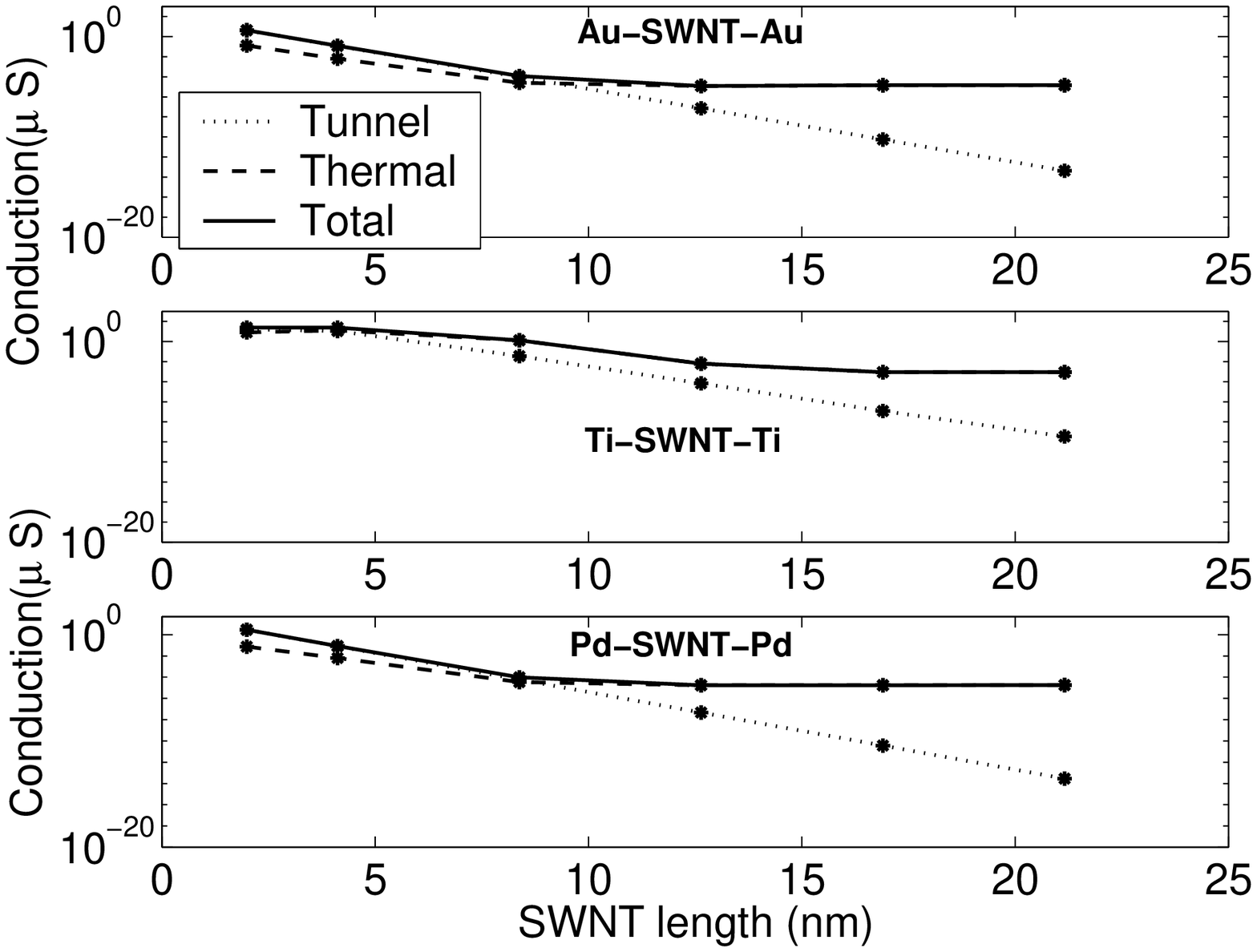}
\caption{\label{xueFig4} Room temperature conductance of the 
metal-SWNT wire-metal junction as a function of SWNT channel length. }  
\end{figure}
          
The SCTB model starts with the semi-emprical Hamiltonian $H_{0}$ of the 
bare $(10,0)$ SWNT wire using the Extended Huckel Theory (EHT) with 
non-orthogonal ($sp$) basis sets $\phi_{m}(\vec r)$,~\cite{Hoffmann88} 
which gives a band gap of $\approx 0.9(eV)$.~\cite{AvouEHT} 
We describe the interaction between the SWNT channel and the rest of 
the junction using matrix self-energy operators and calculate the density 
matrix $\rho_{ij}$ and therefore the electron density of the equilibrium 
SWNT junction from 
\begin{eqnarray}
\label{GE}
G^{R} 
&=& \{ (E+i0^{+})S-H-\Sigma_{L}(E)-\Sigma_{L;NT}(E)
-\Sigma_{R}(E)-\Sigma_{R;NT}(E)\}^{-1}, \\
\rho &=& \int \frac{dE}{2\pi }Imag[G^{R}](E)f(E-E_{F}).
\end{eqnarray}
Here $S$ is overlap matrix and $f(E-E_{F})$ is the Fermi distribution 
in the electrodes. The matrix self-energy operators 
include contributions from coupling both with the left (right) electrodes 
$\Sigma_{L(R)}$ and with the part of the SWNT wire embedded within 
the left (right) electrode $\Sigma_{L(R);NT}$, which are calculated from 
the surface Green's function of the semi-infinite metal surfaces and 
semi-inifinite SWNT wires respectively.~\cite{XueMol,Papa86,Geometry} 
The Hamiltonian of the SWNT channel is 
now $H=H_{0}+\delta V[\delta \rho]$ where $\delta \rho$ is the 
density of transferred charge and $\delta V$ 
is the induced change in the electrostatic potential. 

The self-consistent calculation proceeds by approximating the charge 
distribution as superposition of atom-centered charge 
distributions~\cite{Frau98} $\delta \rho(\vec r)
=\sum_{i} \delta N_{i} \rho_{i}(\vec r-\vec r_{i})$,  
where $\delta N_{i}=(\rho S)_{ii}-N_{i}^{0}$ and $N_{i}^{0}$ is the 
number of valence electrons on atomic-site $i$ of the bare SWNT wire. 
$\rho_{i}(\vec r)$ is a normalized Slater-type function, ~\cite{XueNT,Frau98} 
whose exponent is chosen such that 
$\int d\vec r d\vec r' \rho_{i}(\vec r) \rho_{i}(\vec r')/|\vec r- \vec r'|  
=I_{i}-A_{i}$,~\cite{Frau98} where $I_{i}(A_{i})$ are the atomic electron 
affinity (ionization potential). This gives $\delta V(\vec r)
=\sum_{i} \delta N_{i}V_{i}(\vec r -\vec r_{i})$, where 
$V_{i}=\int d\vec r' \rho_{i}(\vec r'-\vec r_{i})/|\vec r- \vec r'|$ 
can be evaluated analytically.~\cite{XueNT,Frau98} We take into account the 
image-potential effect by including within $\delta V$ contributions from 
both atom-centered charges and their image charges (centered around the 
image positions), rather than imposing an image-type potential correction. 
The self-consistent cycle is completed by calculating the matrix elements 
of the potential $\delta V_{mn}=
\int d\vec r \phi_{m}^{*}(\vec r)\delta V(\vec r)\phi_{n}(\vec r)$  
using two types of scheme: (1) If $m,n$ belong to 
the same atomic site $i$, we calculate it by direct numerical integration; 
(2) if $m,n$ belong to different atomic sites, we use the 
approximation $\delta V_{mn}=S_{mn}(\delta V_{mm}+\delta V_{nn})/2$. 

The calculated charge transfer and electrostatic potential change along 
the cylindrical surface of the SWNT for the Au/Ti/Pd-SWNT-Au/Ti/Pd 
junctions are plotted in Fig. (\ref{xueFig2}). The 
electrostatic potential change is the difference between the electrostatic 
potentials within the SWNT junction and the bare SWNT, obtained 
as the superposition of contributions from the transfered charges 
(plus their image charges) throughout the junction. Here it is 
important to separate the electronic processes at the interface  
and inside the channel. The coupling with electrodes in the 
embedded part of the SWNT wire induces only a localized perturbation to 
the SWNT channel sandwiched between the electrodes, so the electronic 
states in the middle of the channel are essentially identical to 
those of the bulk (infinitely long) SWNT except for the shortest (2.1 nm) 
channel length studied here, leading to similar charge transfer both 
at the interface and in the middle of the channel. For such SWNT channels 
(longer than $ 4.2(nm)$), the potential change in the middle and 
consequently the band lineup scheme become independent of the 
channel lengths. Note that the transfered charge in the middle of the 
channel shows oscillatory behavior due to the two-sublattice 
structure of zigzag tubes.~\cite{XueNT} 

Compared with the finite SWNT molecule contacted to the electrode 
surfaces through the ring of dangling-bond end atoms,~\cite{XueNT} 
the magnitude of charge transfer at the interface is smaller for both 
the Au-SWNT-Au and Ti-SWNT-Ti junctions since no dangling bonds 
are involved in the embedded contact scheme. But the magnitudes of both 
the transfered charge and electrostatic potential change in the middle 
of the SWNT are close to those obtained in the end contact scheme once 
the finite SWNT molecule has reached the bulk limit (20 nm and 
longer).~\cite{XueNT} In addition, although the transfered charge 
and electrostatic potential change are larger at the interface 
for the Pd-SWNT-Pd junction due to the more directional $d$ orbitals 
of Pd surface atoms, their values in the middle of the channel 
are similar to those of the Au-SWNT-Au junction due to the similar 
workfunctions of Au and Pd. For the high (low) workfunction metals as 
Au/Pd (Ti), there is a small decrease (increase) of electron density in the 
SWNT channel, which can be identified as contact-induced hole (electron) 
doping. Therefore, the lineup of the SWNT band in the middle of the 
channel depends strongly on the metal work functions, but only weakly 
on the nature of the contact and the interaction across the interface.  

Le{\'o}nard and Tersoff~\cite{TersoffNT} first pointed out correctly that 
the strength of the interface coupling doesn't affect the lineup of the SWNT 
band in regions moving away from the interface, based on the electrostatics 
of an ideal cylinder and the confined geometry of the metal-nanotube 
interface. However, the electrostatics \emph{of any nanostruture is 
three-dimensional}. For SWNTs the electrostatic potential change induced 
by the charge transfer varies in directions both along and 
perperdicular to the SWNT axis ($Z$ axis). This is clearly seen in 
Fig.\ \ref{xueFig3}(a), where we show a cross sectional view of the 
electrostatic potential change in the $XZ$ plane. For the $(10,0)$ SWNT 
with a diameter of $\approx 0.8(nm)$, the change in the electrostatic 
potential inside such narrow cylinder is small, but decays to about $1/4$ 
of its value at the cylindrical center $1(nm)$ away from the SWNT surface. 

The confined cylindrical geometry and three-dimensional electrostatics of the 
metal-SWNT interface lead to a completely different physical picture 
of the band shift from that of the planar metal-semiconductor interface. 
In particular, the shift of the SWNT band edge in the direction along the 
SWNT axis \emph{doesn't follow the change in the electrostatic potential}. 
This is illustrated in the local density of states (LDOS) of the 
Au-SWNT-Au junction as a function of position along the SWNT axis 
in Fig.\ \ref{xueFig3}(b). Note that despite a $\approx 0.4(V)$ change 
of electrostatic potential within $3(nm)$ of the interface, both the 
conduction band and valence band edges are nearly contant along 
the NT axis.~\cite{LDOS} From the LDOS in the 
middle of the channel, we can determine that for both the Au-SWNT-Au and 
Pd-SWNT-Pd junction the Fermi-level is located slightly below 
(by $\approx 0.05(eV)$) the midgap, while for Ti-SWNT-Ti junction it is 
located above (by $\approx 0.15(eV)$) the midgap. For the case of Au 
and Ti electrodes, these values are essentially identical to those obtained in 
the end contact scheme.~\cite{XueNT}  

The physical principles of Schottky barrier formation at the metal-SWNT 
interface can be understood as follows: Since the electrochemical potential 
(Fermi-level) and therefore the electron occupation in the junction are 
determined by the electrodes in contact with the SWNT,~\cite{Datta} 
the band lineup inside the SWNT channel is determined by the 
self-consistent charge transfer effect through the entire metal-SWNT-metal 
junction. Therefore, the metal Fermi-level position should be close to the 
middle of the gap since otherwise extensive charge transfer will occur 
inside the SWNT channel. Since the screening of the work function 
difference inside the SWNT junction is weak, the metal Fermi-level should 
be below (above) the middle of the gap for a high (low) workfunction 
metal so that the net decrease (increase) of electrons inside the channel 
shifts the SWNT band edge down (up) relative to the metal Fermi-level. 
Exactly how this is achieved from the interface to the middle of the 
channel will depend on the details of the contact without affecting the 
lineup scheme in the bulk region.  

Given the potential shift across the metal-SWNT interface, we can evaluate 
the length and temperature dependence of the SWNT junction conductance 
using the Landauer formula
$G=\frac{2e^{2}}{h}\int dE T(E)[-\frac{df}{dE}(E-E_{F})]=G_{Tu}+G_{Th}$ 
and 
$T(E)=Tr[\Gamma_{L}(E)G^{R}(E)\Gamma_{R}(E)G^{A}(E)]$.~\cite{XueMol,Datta}  
Here we have separated the conductance into tunneling contribution 
$G_{Tu}=\frac{2e^{2}}{h}T(E_{F})$ and thermal-activation 
contribution $G_{Th}=G-G_{Tu}$. The result at room temperature 
is shown in Fig.\  \ref{xueFig4}. The tunneling conductance 
for all three junctions decreases 
exponentially with the SWNT channel length for channel lengths longer than 
$2.1(nm)$. A separation of contact and bulk effect on the tunneling 
resistance can thus be achieved using $R=R_{0}e^{dL}$, 
where $R_{0}$ is the contact resistance and $d$ is the inverse tunneling 
decay length. We find $R_{0}=8.2, 0.68, 12.3 (k\Omega)$ and 
$d=1.68,1.37,1.68(1/nm)$ for the Au-SWNT-Au, Ti-SWNT-Ti 
and Pd-SWNT-Pd junctions respectively. Note that the contact resistance 
is different for the Au and Pd electrodes due to the different interface 
coupling, but the inverse decay length (a bulk-related parameter) 
is the same. The room-temperature conductance saturates with 
increasing SWNT length, because tunneling is exponentially suppressed 
and transport becomes dominated by thermal-activation over the top of 
the potential barrier, whose height is approximately independent 
of the SWNT channel length. For Ti-SWNT-Ti junction, this leads to a 
transition from tunneling to thermally-activated transport at roughly $4(nm)$. 
For Au/Pd-SWNT-Au/Pd junctions, this transition occurs at channel 
length of roughly $9(nm)$. Consequently we can say that transport through 
the SWNT junction is bulk-limited at low temperature, but is contact-limited 
at room temperature.  

In conclusion, we have presented an atomistic real-space analysis of 
Schottky barrier formation at metal-SWNT interfaces with embedded 
contact. Further analysis is needed that treat both the gate and 
source/drain field self-consistently within the SWNT junction, to 
achieve a thorough understanding of nanotube-based devices.   

This work was supported by the DARPA Moletronics program, 
the NASA URETI program and the NSF Nanotechnology Initiative.


\begin{references}
\bibitem{Geometry} We assume the embedded parts of the SWNT are 
surrounded entirely by the metals with overall cylindrical symmetry around 
the SWNT axis. The SWNT-metal interaction arises from one discrete 
cylindrical shell of metal atoms, surrounded by the bulk metal and treated 
using the Green's function method as detailed in Ref.\ \onlinecite{XueMol}. 
We use a SWNT-metal surface distance of $2.0(\AA)$, close to the average 
inter-atomic spacings in the SWNTs and metals.
\bibitem{LDOS} This can be understood as follows: Although the 
charge transfer-induced electrostatic potential change varies 
about $0.4 (eV)$ within $3(nm)$ of the interface, this variation is much 
weaker than the atomic-scale electrostatic potential variations within the 
bare SWNT. Its effect on the LDOS is therefore similar to that of 
small molecules in the molecular tunnel junction, where detailed studies in 
Ref. \onlinecite{XueMol} have shown the LDOS doesn't change much with 
position despite a $\sim 1(eV)$ change in electrostatic 
potential within $\sim 1(nm)$ in the molecular junction.
\end{references}
\end{document}